\documentclass[aps,showpacs,preprint,a4paper,12pt]{revtex4}
\usepackage{graphicx,color}
\usepackage{subfigure}
\usepackage{amssymb}
\usepackage{hyperref}
\usepackage{srcltx}
\begin{document}
\title{Asymmetric Exclusion Process in a System of
Interacting Brownian Particles}
\author{Jos\'e Eduardo de Oliveira Rodrigues}
\email[E-mail: ]{jeo@fisica.ufmg.br}
\author{Ronald Dickman}
\email[E-mail: ]{dickman@fisica.ufmg.br}
\affiliation{Departamento de F\'isica, ICEx, Universidade Federal de Minas Gerais, Caixa Postal 702,
Belo Horizonte 30161-970, Minas Gerais, Brazil}
\begin{abstract}
We study a continuous-space version of the totally asymmetric simple exclusion
process (TASEP), consisting of interacting Brownian particles subject to a
driving force in a periodic external potential. Particles are inserted at the
leftmost site at rate $\alpha$, hop
to the right at unit rate, and are removed
at the rightmost site at rate $\beta$. Our study is motivated by recent
experiments on colloidal particles in optical tweezer arrays.
The external potential
is of the form generated by such an array. Particles spend
most of the time near potential minima, approximating the situation in the
lattice gas; a short-range repulsive interaction prevents two particles from
occupying the same potential well. A constant driving force, representing
Stokes drag on particles suspended in a moving fluid, leads to biased motion.
Our results for the density profile and current, obtained via
numerical integration of the
Langevin equation and dynamic Monte Carlo simulations,
indicate that the continuous-space model exhibits
phase transitions analogous to those
observed in the lattice model.  The correspondence is not exact, however, due to
the lack of particle-hole symmetry in our model.
\end{abstract}
\maketitle

\section{Introduction}

A driven lattice gas, or driven diffusive system,
is a system of interacting particles that jump in a preferred direction on
a lattice.
The system cannot reach equilibrium but may attain a stationary
state with a steady current;
the model is a prototype for studies of nonequilibrium states
\cite{Schmittmann,Marro,Katz,Leung}. The simplest example of a driven
diffusive system, which has become one of the standard models of nonequilibrium statistical
mechanics, is the totally asymmetric simple exclusion process (TASEP)
\cite{MacDonald-Gibbs2,Andjel,Harris,Sptizer}.
In the TASEP with open boundaries
the edge sites are connected to particle reservoirs with fixed densities.
Introduced as a model of biopolymerization
\cite{MacDonald-Gibbs} and transport across membranes \cite{Heckmann}, over the years, this
model has been applied to other processes, e.g., traffic flow
\cite{Schadschneider}, and cellular transport \cite{Heijne,Kukla}.

From a mathematical point of view the model is of interest in the theory of interacting
particle systems since, despite its simplicity, it shows a nontrivial
behavior \cite{Harris,Sptizer,Ligget85,Spohn}. In the one-dimensional TASEP with open
boundaries, particles jump only to the right, along a
one-dimensional lattice whose sites can be empty or occupied by a single particle.
Particles are injected at the leftmost site
at rate $\alpha$ if this site is empty, and removed at the rightmost site
at rate $\beta$ if this site is occupied. The one-dimensional TASEP, which has been solved
exactly, exhibits three distinct
phases in the $\alpha$-$\beta$ plane
\cite{SchutzDomany,Derrida93i,Derrida96,Derrida98}.
The phase transition is discontinuous along the
line $\alpha=\beta<1/2$, where the density profile is linear, and continuous along the
lines $\alpha=1/2$, $\alpha>\beta$ and $\beta=1/2$, $\alpha<\beta$ (see Fig.~\ref{PhaDia}).
Although this model and and variants have been the subject of intensive theoretical
study, there is as yet no realization of a TASEP-like system in the laboratory.

The invention of optical tweezer arrays has permitted investigation of
the dynamics of colloidal particles in an external periodic potential
\cite{Sancho,Korda,cellsorting,Chiou,Lacasta,Roichman}. The motion at
long times and low friction consists of jumps between adjacent
potential minima. If the particle and potential well sizes are chosen properly, only
one particle can occupy a given well.  The exclusion process is a caricature of this dynamics,
suggesting that a system of colloidal
particles in an optical tweezer array could be designed as a laboratory realization of the TASEP.
Motivated by this possibility, we propose a model in continuous space having the
same essential characteristics as the lattice TASEP.
Our model represents colloidal particles immersed in a fluid flowing at constant rate
through a one-dimensional optical tweezer array, restricting particle motion to the array axis.
Specifically, we study a one-dimensional system of interacting Brownian particles
subject to a periodic external potential,
and to a constant external force representing the drag due to
the fluid motion.  A short-range (essentially hard-sphere) repulsion between particles
prevents more than one particle occupying the same potential well.
This continuous-space model is studied via numerical integration of Langevin equation
and dynamic Monte Carlo simulation.
We observe phase transitions similar to those
found in the lattice TASEP.  Some differences in the detailed behavior nevertheless appear,
due to the lack of particle-hole symmetry in the continuous-space model.
Details on the model and simulation methods are given in Sec.~\ref{LAN} and \ref{MC}.
Simulation results are presented in Sec.~\ref{RES},
while our conclusions and prospects for future work are outlined in Sec.~\ref{CON}.

\section{Continuous-Space Model}\label{LAN}

Our aim is to study a continuous-space model sharing the same essential features
as the TASEP (defined on a lattice), as a first step toward
experimental realization of a TASEP-like system.
The model should possess the following characteristics: i) confinement of particles
to a one-dimensional structure; ii) localization of particles at potential minima (``wells") of a
linear periodic array with iii) multiple occupancy prohibited; iv) biased hopping between
adjacent wells; v) insertion (removal) of particles at the initial (final) well.
Criteria  i)-iii) are realizable with a suitably tailored optical tweezer array.
The array consists of a series of spherically symmetric optical tweezers;
the particles flow along this line, which we take as the $x$ axis.
To avoid particles escaping the array,
there should be a substantial overlap between neighboring wells,
so that a potential maximum
at a point midway between two wells is in fact a saddle point in the full three-dimensional
space (see Fig.~\ref{two_gau_inv}).
For a TASEP-like
system, it is crucial that the probability of a particle escape from the array be
negligible on the time-scale of the experiment.
Effective confinement requires a potential barrier
to escape the array much larger than $k_BT$; the barrier
between adjacent minima should be smaller, to
allow transitions between
neighboring wells.
In what follows we shall assume this condition is satisfied, and consider,
for simplicity, a one-dimensional system.  Fluctuations of the particle positions
in the directions perpendicular to the array will therefore be ignored, but should be
included in a more complete analysis.

The diameter of the optical tweezer well should be slightly greater than the
particle diameter, so that at most one particle can occupy the well at a given time.
Due to thermal fluctuations,
particles can occasionally overcome the potential barrier
separating neighboring wells.  Since hopping must be
asymmetric, we impose a
steady fluid motion along the $+x$ axis, which effectively prohibits particle jumps in
the opposite direction.
In the lattice TASEP particles are inserted in the first site and removed from the last.
Experimental realization of this feature is subtler, but can in principle be achieved
with the help of
optical tweezers at the beginning and end of the array, which drag particles
into the first well and out of the last one at prescribed rates.  We discuss
an alternative method of insertion and removal in Sec.~\ref{CON}.

The above sketch of an experimental setup motivates our study
of Brownian motion of interacting colloidal
particles in an optical tweezer array.
The Langevin equation for the $i$-$th$ particle is

\begin{equation}\label{ELUP}
m\ddot{x}_i(t) = -b \left[ \dot{x}_i(t) - v\right]  - \frac{\partial V_{ext}(x_i)}{\partial x_i}
- \frac{\partial}{\partial x_i} \left[ V_{int}(x_{i,i-1}) + V_{int}(x_{i,i+1})\right] + m\Gamma_i(t),
\end{equation}

\noindent where $x_i$, $\dot{x}_i$ and $\ddot{x}_i$ are, respectively, the position, velocity and acceleration
of particle~$i$, $x_{i,j} \equiv x_j - x_i$, and $v$ is the terminal velocity of a particle
in the moving fluid, in the absence of the periodic external potential $V_{ext}$.
$V_{int}$ is the (strongly repulsive) potential between neighboring
particles.
(We assume that the range of $V_{int}$ to be short enough that only neighboring particles interact.)
The first term on the right side of Eq.~(\ref{ELUP}) represents damping of the particle
velocity relative to the fluid. For a sphere of radius $R$, Stokes' Law gives:
\begin{equation}\label{EscStokes}
b = 6\pi\mu R,
\end{equation}
where $\mu$ is the fluid viscosity.
Here it is important
to stress that the fluid is three dimensional although we treat the particle motion
as one dimensional. The final term is a random noise
with the following properties:
\begin{eqnarray}\label{PRA}
 \langle\Gamma_i\rangle &=& 0, \\
 \langle\Gamma_i(t)\Gamma_j(t')\rangle &=& \frac{2b k_B T}{m^2}\delta_{i,j}\delta(t-t'),
\end{eqnarray}

\noindent where $k_B = 1.3806504\times10^{-23}J/K$ is Boltzmann's constant, $T$ is temperature,
and $m$ is the particle mass.
It is convenient
to ignore the inertial term $m\ddot{x}_i(t)$ in Eq.~(\ref{ELUP}),
since the observational times of interest (microseconds or greater) are
much larger than the
relaxation time of the velocity, $m/b\sim10^{-8}\,$s, for our choice of parameters.
Then the velocity of particle $i$ follows,
\begin{eqnarray}\label{ELUP2}
\dot{x}_i(t) = v  - \frac{1}{b}\frac{\partial V_{ext}(x_i)}{\partial x_i} -
 \frac{1}{b}\frac{\partial}{\partial x_i} \left[ V_{int}(x_{i,i-1}) + V_{int}(x_{i,i+1})\right]
+ \frac{m}{b}\Gamma_i(t).
\end{eqnarray}

\subsection{Potentials}

\begin{figure}[htb]
\begin{center}
\subfigure[\label{two_gau_inv}]{\includegraphics[height = 4.6cm]{two_gauss}}
\subfigure[\label{pot_efe}]{\includegraphics[height = 4.6cm]{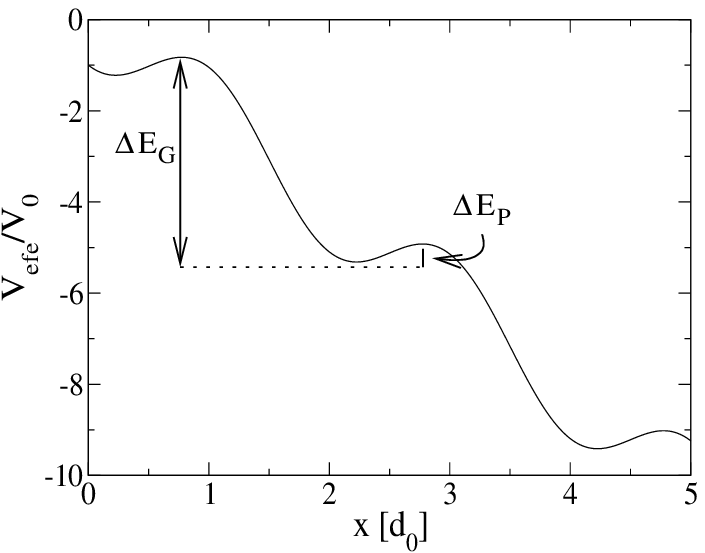}}
\end{center}
\caption{(a) Two neighboring Gaussian profiles (red and black curves) and their sum (green);
(b) effective external potential.}
\end{figure}

The potential of an optical tweezer array can be represented by a sum of $n$ identical
Gaussian profiles of width $\sigma$ and spatial period $d$:
\begin{equation}
V_t (x) = - V_0 \sum_{n=0}^N e^{-(x-dn)^2/2\sigma^2}.
\end{equation}
We require neighboring wells to overlap, which can be accomplished
by setting $\sigma \stackrel < \sim d$; Figure~\ref{two_gau_inv} shows that
this condition is satisfied for
$\sigma=d/4$.
For this choice of parameters, the
potential is well approximated by a cosine, as can be seen from the
Fourier coefficients
\begin{equation}\label{CoeFou}
 a_n = \frac{2}{d}\int\limits_0^d\cos\left(\frac{2n\pi x}{d}\right)
\left[ -\sum\limits_{n=0}^N e^{-(x-dn)^2/2\sigma^2}\right] dx.
\end{equation}

\noindent Numerical evaluation yields

\begin{equation}\label{CoeFouNum}
\begin{array}{l}
 a_0 = -0.627\\
 a_1 = 0.365\\
 a_2 = 9.010\times10^{-3}.
\end{array}
\end{equation}
Note that $a_2/a_1 \simeq 0.025$, allowing us to write,
to a good approximation, the array potential as a constant plus a
cosine term.  For the one-dimensional model studied here, we define
\begin{equation}\label{EnePotExt}
 V_{ext}(x) = -V_0 \cos\left(\pi \frac{x}{d_0}\right)
\end{equation}
as the periodic external potential.

The interaction between
colloidal particles is taken as purely repulsive; for convenience we use a truncated
$1/r^{12}$ potential:
\begin{equation}\label{EnePotInt}
 V_{int}(r) = \left\lbrace \begin{array}{cc}
 \displaystyle U_0\left[\left( \frac{a}{r}\right) ^{12} -1\right] , & r\leq a\\
 0, & r\geq a,
\end{array}\right.
\end{equation}
where $a$ is the particle diameter and $r$ distance between neighboring particles.

\subsection{Parameter Values}

To specify the external and interaction potentials, we need to fix
$V_0$ and $U_0$. These values must be chosen so as to approximate the TASEP
dynamics given the length, time and energy scales characterizing the system.
We measure lengths in units of microns, time in seconds and energies in units of
$k_BT =  4.1419512\times10^{-21} $J, assuming a temperature of 300K.
We set $d_0=1$ (so that the period of the external potential is $2$ microns),
and take the particle diameter as $a=1.8$, so that a pair of particles occupying neighboring wells have
some freedom to fluctuate about the potential minimum. Taking the fluid as water
(with viscosity $\mu = 0.01$g/cm$\cdot$s) we fix the friction coefficient as
$b\cong4.096\;k_BT\cdot $s/($\mu$m)$^2$.

To determine the fluid velocity $v$ and external-potential intensity $V_0$ we examine
the effective external potential, defined as
\begin{equation}\label{PotExtEfe}
V_{eff} = -V_0\cos\left(\pi \frac{x}{d_0}\right)- b v x,
\end{equation}

\noindent i.e., the sum of external periodic potential and a fictitious potential representing
the constant friction force acting on particles (see Fig.~\ref{pot_efe}). If $b v d_0 \gg V_0$, particles
do not feel the periodic potential, while if $b v d_0 \ll V_0$ particle
hopping is essentially unbiased. Let $\Delta E_G$ ($\Delta E_P$)
denote the difference between a given maximum of $V_{eff}$ and the first minimum to
the right (left) of this maximum. In this way, $\Delta E_G$ ($\Delta E_P$) is the
potential barrier separating a given potential minimum from its left (right)
neighbor. A simple calculation shows that
\begin{equation}\label{DeltaE}
\begin{array}{l}
\Delta E_G - \Delta E_P = 2 b v d_0 \,,\\
\Delta E_G + \Delta E_P = 4V_0\cos(\pi x_0/d_0) + 4b v x_0 \,,
\end{array}
\end{equation}
with
\begin{equation}\label{x0}
 x_0 = \frac{d_0}{\pi} \sin^{-1}\left( \frac{b v d_0}{\pi V_0}\right)\,,
\end{equation}
such that the potential minima occur at $x_j = 2jd_0+x_0$, and maxima at $y_j = 2jd_0+1-x_0$ for
$j$ an integer.
Given $\Delta E_G$ and $\Delta E_P$, we can determine $x_0$,
$v$ and $V_0$. We take $\Delta E_P \sim k_BT$, and $\Delta E_G \gg k_B T$, so that
a particle has a finite rate of jumping to the well on the right, and virtually
no chance of jumping in the opposite direction.
A good correspondence with lattice TASEP is obtained using
$\Delta E_G = 20k_BT$ and $\Delta E_P = 2k_BT$.
For these values we have,
\begin{equation}\label{V_0vx_0}
\begin{array}{l}
 v = 2.1974\;\mu\mbox{m/s}\\
 V_0 = 4.5680 \;k_BT\\
x_0 = 0.2158 \;\mu\mbox{m}.
\end{array}
\end{equation}

\noindent To maintain the interparticle  repulsion in the presence of the
external potential, we must take $U_0$ substantially greater than $V_0$.
On the other hand, very large values of $U_0$ are inconvenient for
numerical integration of the Langevin equation, as a very small time increment
would be required to avoid spurious particle displacements.
We therefore use $U_0 = 40k_BT \simeq 9 V_0$.

\subsection{Single-particle dynamics}

\begin{figure}[htb]
\begin{center}
\includegraphics[height=6cm]{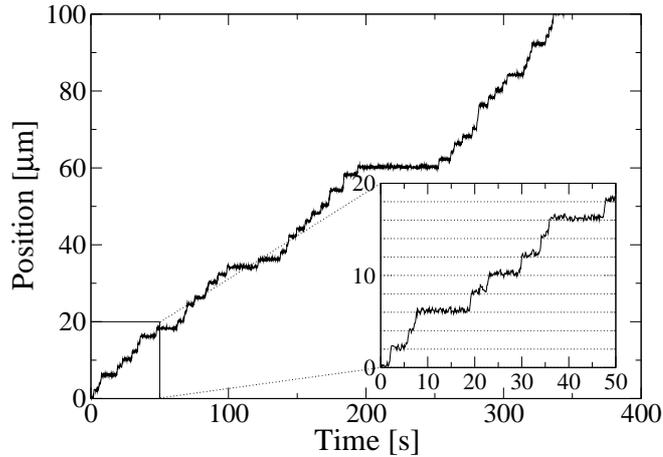}
\end{center}
\caption{Typical particle trajectory.}
\label{position-time}
\end{figure}

To begin, we consider a single Brownian particle moving in the
system. Figure~\ref{position-time} shows a typical evolution of the particle position
over time. The graph exhibits plateaux whose size corresponds
to the time a particle stays in a given well. Studies of this kind allow us to determine
the mean transition time $\tau$ between neighboring wells as 6.5380(9)s. This quantity is needed in order to
define the insertion and removal rates.  (Recall that in
the lattice TASEP these rates are defined in units of the hopping rate).
If the first well is empty
we insert a particle there (at the potential minimum position), at rate $\alpha/\tau$;
if the last well is occupied, we remove
the particle at rate $\beta/\tau$.

\section{Dynamic Monte Carlo Simulations}\label{MC}

The Langevin simulation outlined in the preceding section is valuable for fixing the
time scale of hopping between wells, and for confirming the basic phenomenology of the
model.  It is, however, rather inefficient numerically, so that it is desirable to implement
a dynamic Monte Carlo (MC) simulation for large-scale studies.
We apply the Metropolis algorithm to evolve the particle positions
in time.

In this approach we use the following expression for the potential energy:

\begin{equation}\label{EneTotSis}
E = -\sum_{i=1}^N [V_0\cos(\pi x_i) + b v x_i ] + \sum_{i=1}^{N-1} V_{int}(x_{i+1}-x_i) ,
\end{equation}
where $V_{int}$ is given by Eq.~(\ref{EnePotInt}). The drag force due to the moving fluid
is represented by the effective potential $-b v x$.
In the MC dynamics,
a trial configuration is generated by selecting one of the $N$ particles at random
and subjecting it to a random displacement $\Delta x$, chosen from
a Gaussian distribution with mean zero and standard deviation $\sigma~=~0.2\mu$m.
This value is $10\%$ of the well size,
large enough to afford a substantial speedup, but small enough that the probability of a
particle displacement greater than $2\mu$m is negligible.
As is usual in Metropolis MC, trial moves such that the change in energy
$\Delta E \leq 0$ are always accepted,
while for $\Delta E > 0$ the trial move is accepted with probability $e^{-\Delta E/k_B T}$.
We determine the mean
number of Monte Carlo steps required for a particle move from one well to its
neighbor on the right as
$N_{MC} = 120.917(1)$.
Thus the time per Monte Carlo step is
\begin{equation}\label{tc}
 \tau_0 = \frac{\tau}{N_{MC}} = 5.4070(8)\times10^{-2}\mbox{s}.
\end{equation}

\noindent Particle insertion and removal are done as in the Langevin simulations.
We verify below
that this method is equivalent to the latter approach; the MC algorithm is
$\sim \! 1000$ faster than numerical integration of the Langevin equation,
for L=100.
\vspace{1cm}

\begin{figure}[htb]
\begin{center}
\includegraphics[height=6cm]{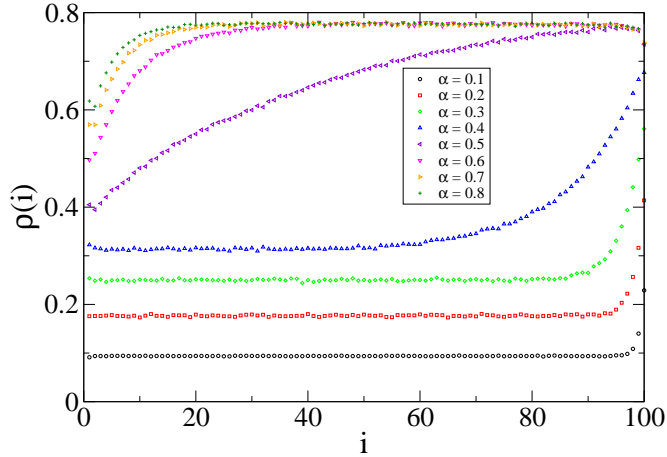}
\end{center}
\caption{Density profile for $\beta = 0.4$ obtained via numerical integration of the
Langevin equation, for $\alpha$ values
as indicated, $L=100$.}
\label{profile}
\end{figure}

\section{Results}\label{RES}

\begin{figure}[htb]
\begin{center}
\includegraphics[height=6cm]{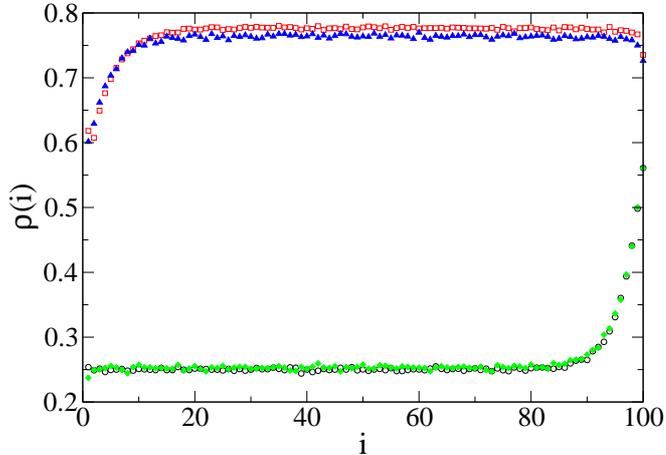}
\end{center}
\caption{Comparison between profiles obtained using LS (open symbols) and
MC (filled symbols), $\alpha=0.3$ (lower curves), $\alpha=0.8$ (upper curves), $\beta=0.4$ , $L=100$.}
\label{MCLAComp}
\end{figure}

During the simulations we monitor the mean occupation probability $\rho(i)$ at each well $i$,
and the current $J$,
measured by the mean number of particles leaving the system per unit time.
Examples of density profiles in the stationary regime
($\rho(i)$ versus $i$), are shown in
Fig.~\ref{profile} for several values
of $\alpha$, with $\beta = 0.4$.
(From here on $\alpha$ and $\beta$ are given in units of $1/\tau$, where $\tau$ is the
mean time required for hopping between wells.)
This figure shows that the continuous-space model exhibits the same basic phenomenology
as the lattice TASEP.  For
$\alpha \leq 0.4$ the overall density grows with $\alpha$.  On increasing
$\alpha$ from 0.4 to 0.5 there is a marked increase in density, but for further increases
the density changes very little.
While the Langevin simulation (LS) results already suggest that the model exhibits
phase transitions, we shall use the more precise results
of our MC simulations to perform a detailed analysis.  Before proceeding, we
verify that the MC method yields results in agreement with the LS. In Fig.~\ref{MCLAComp}
we compare density profiles obtained via
LS and MC for the same values of $\alpha$ and $\beta$.  The bulk densities
obtained using the two methods differ by $\le 1.6 \%$.  Thus the MC method captures the behavior
found using the
Langevin equation to good precision.

\begin{figure}[htb]
\begin{center}
\includegraphics[height=6cm]{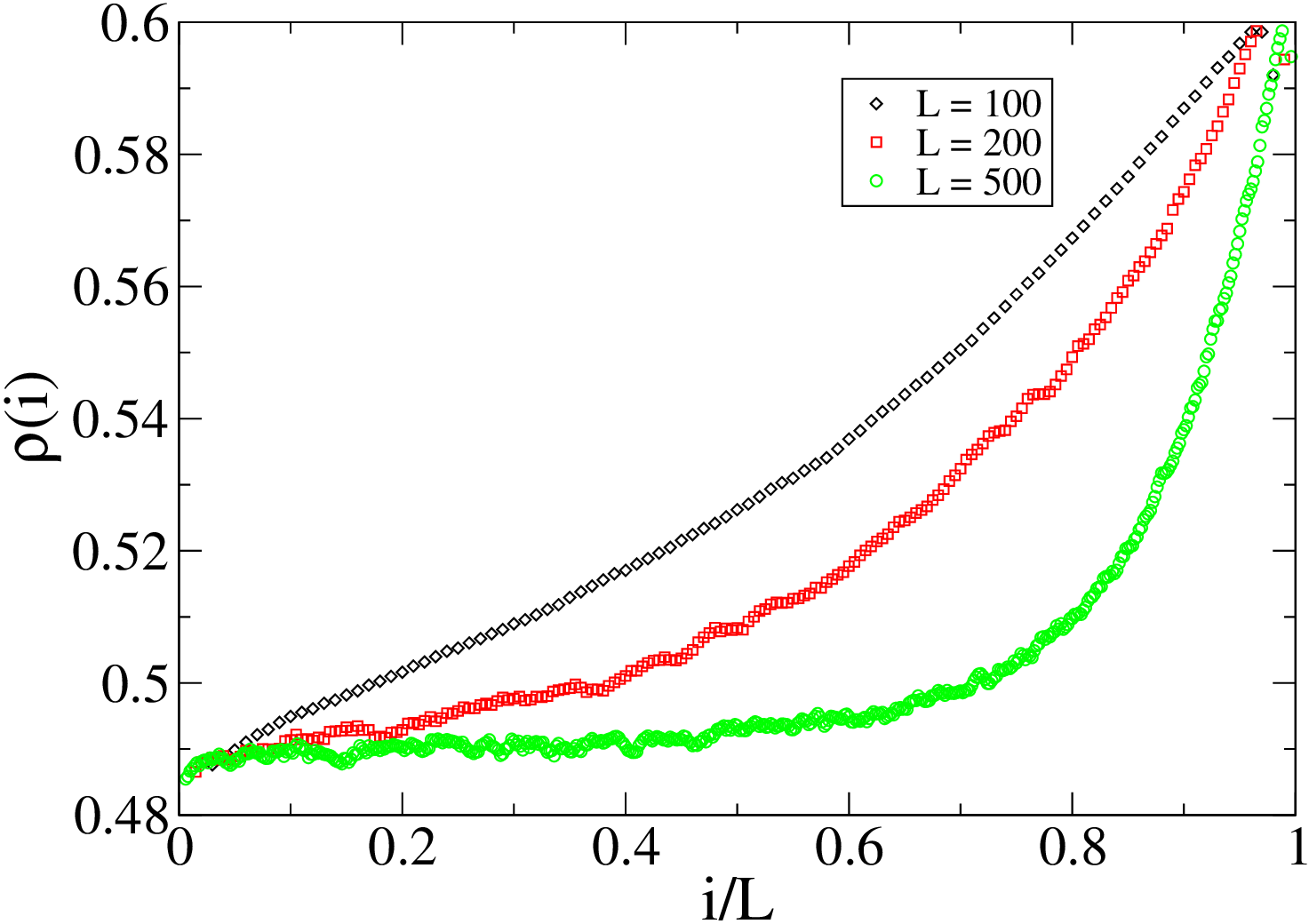}
\end{center}
\caption{Density profiles for $\alpha = 0.7$ and $\beta = 0.6$, for systems of
$100$, $200$, and $500$ wells.} \label{SizPro}
\end{figure}

We perform MC simulations of systems of $L=100$, 200 and 500 wells.
Far from the phase transition, density profiles depend only weakly
on system size, but near the transition there are significant
finite-size effects, as illustrated in Fig.~\ref{SizPro}. In this case, as $L$
increases, the profile tends to a near-constant value except for a
sharp increase near the exit.
\vspace{2em}

To minimize boundary effects we study the {\it bulk} density
$\rho$, defined as the mean density over the $10\%$ of sites nearest
the center:

\begin{equation}\label{density}
 \rho = \frac{10}{L}\sum_{i=0.46L}^{0.55L}\rho(i).
\end{equation}

\begin{figure}[htb]
\begin{center}
\includegraphics[height=6cm]{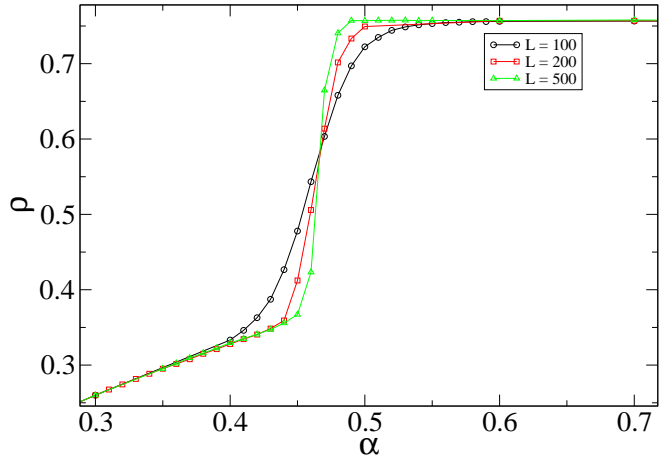}
\end{center}
\caption{Bulk density versus $\alpha$ for $\beta = 0.4$, system
sizes $L=100$, $200$ and $500$. } \label{SysDenDifSiz}
\end{figure}

\noindent The bulk density as a function of $\alpha$, for $\beta = 0.4$, is
shown in Fig.~\ref{SysDenDifSiz} for the three system sizes studied.  These
results strongly suggest the development of a discontinuity in
$\rho(\alpha)$ near $\alpha = 0.47$ as the system size is increased.

\begin{figure}[htb]
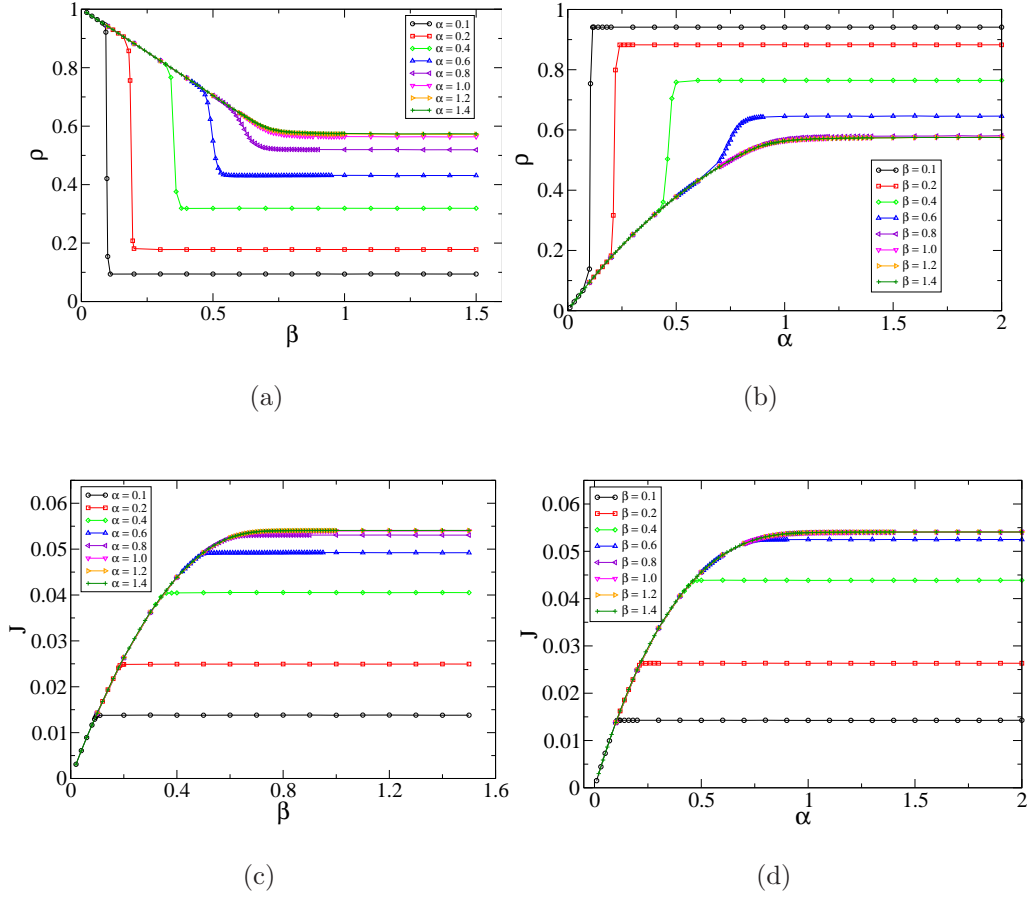

\center
\subfigure[\label{MosImpRes_a}]{\includegraphics[height = 4.6cm]{DenAlfFix2}}
\subfigure[\label{MosImpRes_b}]{\includegraphics[height = 4.6cm]{DenBetFix2}}
\\ \vspace{0.5cm}
\subfigure[\label{MosImpRes_c}]{\includegraphics[height = 4.6cm]{CorAlfFix2}}
\subfigure[\label{MosImpRes_d}]{\includegraphics[height = 4.6cm]{CorBetFix2}}
\caption
{Upper panels: bulk density for (a) fixed $\alpha$ and (b) fixed $\beta$. Lower panels:
current for (c) fixed $\alpha$ and (d) fixed $\beta$. System size $L = 200$.
}
\label{MosImpRes}
\end{figure}

\subsection{Phase diagram}

Of principal interest in determining the phase diagram
are the bulk density and the current as functions of the
rates $\alpha$ and $\beta$.  These results are summarized in
Fig.~\ref{MosImpRes}, showing evidence of both continuous and discontinuous phase
transitions, depending on the rates.
We see that for low $\alpha$ ($\alpha<0.8$ or so) and $\beta<\alpha$
the system is in the high-density phase, in which  density and current depend only on $\beta$,
whereas for $\beta > \alpha$, and $\beta < 0.6$ or so, the system is in the low-density phase in which
density and current depend only on $\alpha$.  For larger values
($\alpha>0.8$ and $\beta>0.6$), the system is in the maximum-current phase, in which density and
current are independent of both $\alpha$ and $\beta$, and the current takes its maximum value.
Thus the continuous-space model exhibits the same
three phases observed in lattice model. As in the lattice model, the transition
between the low- and high-density phases is discontinuous, whereas transitions between
the maximum-current phase and the other phases are continuous.
(While the density is discontinuous in the former case, the current is always
continuous at the transition.)

We adopted the following procedure (``polynomial method") to determine the values of
$(\alpha,\beta)$ along the discontinuous
transition line.  Consider the case of fixed $\alpha$.
For small $\beta$, the stationary current depends only on $\beta$ (see Fig.~\ref{MosImpRes_c}).
We therefore fit a polynomial $P(\beta)$ to the current in this regime, using data for
a large, fixed $\alpha$ (in practice, $\alpha \geq 1.4$).
For larger values of $\beta$, the current depends only on $\alpha$;
in Fig.~\ref{MosImpRes_c} this regime corresponds to one of the plateaux, $J = J^* (\alpha)$.  The transition
point $\beta_c (\alpha)$ is taken as the value at which the plateau intersects the polynomial
fit to the small-$\beta$ data, i.e., $P(\beta_c) = J^* (\alpha)$.
Determination of $\alpha_c (\beta)$ follows an analogous procedure,
in which we fit a polynomial to the current data for small $\alpha$.
We find that a quadratic polynomial is sufficient to fit (to within uncertainty) the current
$j(\beta)$ at fixed $\alpha$, while a good fit of $j(\alpha)$ (at fixed $\beta$) requires a
quartic polynomial.  In both cases the constant term in the polynomial is zero, since
the current vanishes for $\alpha$ and/or $\beta$ zero.

\begin{figure}[htb]
\begin{center}
\includegraphics[height=6cm]{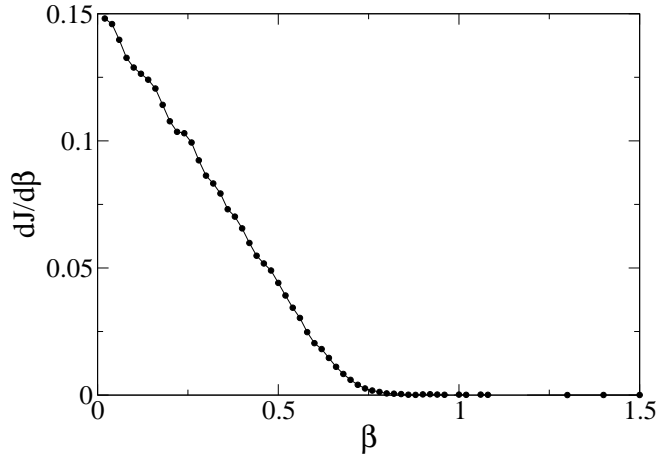}
\end{center}
\caption{Current derivative for $\alpha=2.0$ and $L=200$.}
\label{DC}
\end{figure}

The above method is quite effective in locating points along the discontinuous transition line.
Although it can in principle be used to locate continuous transitions as well,
we found that in this case the estimates for $\alpha_c$ and $\beta_c$
are rather sensitive to one's choice of the range of values fit using the polynomial.
We found the following
approach (``derivative method") to be more useful for continuous transition points.
For fixed $\alpha$, we estimate the derivative $dJ/d\beta$ using a spline fit
(see Fig.~\ref{DC}).  The derivative decreases in a linear fashion with increasing $\beta$,
except for a small roundoff region that we interpret as a finite-size effect.  We then estimate
$\beta_c$ as the point where $d J/d \beta$ falls to zero, using the data in the linear region.
The procedure for fixed $\beta$ is analogous.
The continuous transition points obtained by this procedure are quite robust with
respect to changes in the region analyzed, as long as we exclude the roundoff region.

\begin{figure}[htb]
\begin{center}
\includegraphics[height=6cm]{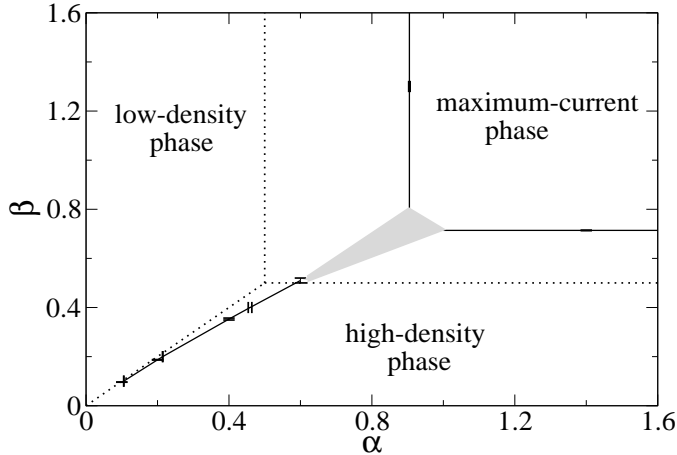}
\end{center}
\caption{Phase diagram of the TASEP on the lattice (dotted lines) and
in continuous-space (solid lines)
for L=500. The grey triangle denotes a region in which we are unable to
determine the phase boundaries precisely.}
\label{PhaDia}
\end{figure}

Using the methods described above we construct a phase diagram based on the data for each system size
studied; that for $L=500$ is shown in
Fig.~\ref{PhaDia}. The vertical boundary
at $\alpha=0.905(2)$, and the horizontal boundary at $\beta=0.714(1)$ are obtained
using the derivative method.  (The polynomial method yields $\alpha = 0.8(1)$ and $\beta = 0.67(5)$,
respectively, for these boundaries.)
The grey triangle in Fig.~\ref{PhaDia} represents a region on which we could not determine precisely
the phase boundaries using either method, due to numerical uncertainty and finite-size effects.
While the simplest interpretation is that the three phase boundaries meet at the point of intersection between
the continuous transitions (i.e., horizontal and vertical) lines, the data in hand are not sufficient
to verify this.

The phase boundaries for $L=100$ and $L=200$ are nearly the same as for $L=500$,
suggesting that the latter
are already quite close to limiting (infinite-$L$) values.
Table~\ref{CTP} gives the $\alpha$ and $\beta$ values for the continuous transition
lines for the three sizes.  There is little sign of systematic variation with system size,
other than a small ($\sim 2.5$\%) increase in $\beta_c$ on going from $L=200$ to $L=500$.
\begin{table}[htb]
\begin{center}
  \begin{tabular}{ c  c  c }
    \hline
    $L$ & $\alpha_c$ & $\beta_c$ \\ \hline \hline
    100 & 0.901(4)   & 0.701(2) \\
    200 & 0.901(2)   & 0.697(1) \\
    500 & 0.905(2)   & 0.714(1) \\
    \hline
  \end{tabular}
\end{center}
\caption{Values of $\alpha$ and $\beta$ at the continuous transitions
obtained via the derivative method.}
\label{CTP}
\end{table}

While the continuous-space phase diagram is isomorphic to that of the lattice model,
there are some differences between the two cases that appear
likely to persist in the infinite-size limit.
Since the lattice model possesses particle-hole symmetry, the phase diagram is
invariant under the exchange of $\alpha$ and $\beta$.  Thus the boundary
between the high- and low-density phases is a straight line extending from the
origin to the point $(1/2,1/2)$ in the $\alpha$-$\beta$ plane.
The phase diagram of the continuous-space model does not possess this symmetry;
the phase boundary between the high- and low-density
phases does not fall along the line $\alpha=\beta$, and appears to be somewhat curved.

One might inquire whether the differences in the phase boundaries of the
lattice and continuous-space models merely reflect finite-size effects
in the latter.  We have verified that in the {\it lattice} TASEP with $L=200$, the phase boundaries
fall quite near their expected (infinite-size) positions.  Comparison of the
phase boundaries (in continuous space) for $L=200$ and 500 suggests that
finite-size effects are somewhat stronger in the continuous space model than on the lattice.
Given the lack of particle-hole symmetry, however, it appears very unlikely that the
continuous-space phase boundaries will converge to those of the lattice model
in the infinite-size limit.

The differences between the lattice and continuous-space models reflect, in part, the absence
of particle-hole symmetry in the latter; particle positions fluctuate in continuous space,
but are fixed in the lattice model.
In continuous-space, moreover, particles occupying neighboring wells may influence one another
via the repulsive potential $V_{int}$. On the lattice model no such influence exists,
beyond simple exclusion.  In continuous space,
repulsive interactions should tend to spread particles
more uniformly than on the lattice, promoting particle removal, and hindering insertion.
Thus the transition from high to low density occurs for $\beta < \alpha$.
Since repulsion is more significant for
higher densities (i.e., larger $\alpha$) the phase boundary should curve toward the $\alpha$
axis, as is observed.  The smaller value of $\beta$ at the continuous transition (approximately
0.714(1) for $L=500$), compared to that of $\alpha$
(about 0.905(2) for $L=500$) may also be attributed
to repulsion between neighboring particles.

\subsection{Current: comparison with mean-field theory}

\begin{figure}[htb]
\begin{center}
\includegraphics[height=6cm]{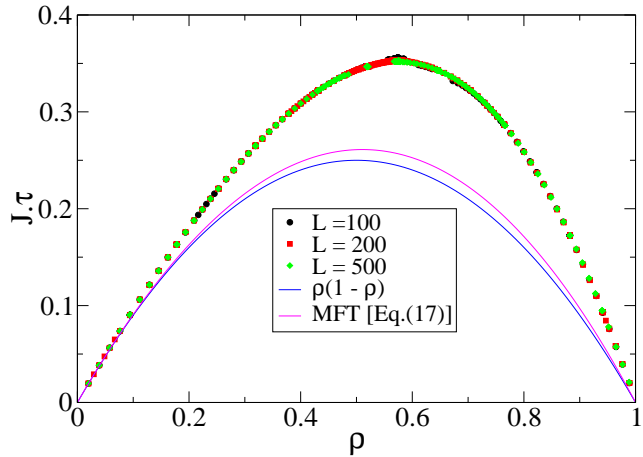}
\end{center}
\caption{Current \textit{versus} density in the lattice model (exact) and the continuous-space model.}
\label{J_d}
\end{figure}

The absence of particle-hole symmetry is again evident in a plot of the current as a function
of density.
In the lattice model, mean field theory gives $J = \rho(1-\rho)$ \cite{Krug91,Derrida92}
which is in fact an exact expression.
Fig.~\ref{J_d} compares the current on
lattice with our results for continuous space. (Note that the latter
exhibit virtually no finite-size effects on the scale of the figure.)
Unlike in the lattice TASEP, here the current is not
symmetric about $\rho=1/2$; it takes
its maximum value at a density of about 0.57.
The fact that the maximum current occurs at a higher density than on the lattice
may again be attributed to interparticle repulsion.
\vspace{2em}

On the lattice, the current is equal to the
probability of having an occupied site with its neighbor on the right vacant, i.~e.,
$J~=~\langle\xi_i(1-\xi_{i+1})\rangle, \; \forall i$, where $\xi_i$ is an indicator variable
equal to one if site $i$ is occupied and $0$ if it is empty.
In mean-field theory the joint probability is factored so:
$\langle\xi_i (1-\xi_{i+1})\rangle = \langle\xi_i\rangle\langle(1-\xi_{i+1})\rangle$,
and setting $\langle\xi_i\rangle = \rho$ in the bulk, we obtain
$J = \rho(1-\rho)$.

In developing a mean-field theory for the continuous-space model,
one might argue that $J \simeq \rho(1-\rho)/\tau$, i.~e., that the current is simply
given by the transition rate $1/\tau$ for jumps between neighboring wells times the
probability of a given well being occupied and its right neighbor empty.
Due to repulsive interactions
between neighboring particles, however, a particle in the well to the left can also influence
the hopping rate. As a first approximation we write
\begin{equation}\label{MFTC}
 J = P(0,1,0)j(0,1,0) + P(1,1,0)j(1,1,0),
\end{equation}
where $P(\xi_{i-1},\xi_i,\xi_{i+1})$ is the joint probability for
three adjacent wells, and $j(\xi_{i-1},\xi_i,\xi_{i+1})$
is the transition rate in this configuration. Factorizing the joint probability,
we have $P(0,1,0) = \rho (1-\rho)^2$ and $P(1,1,0) = \rho^2 (1-\rho)$. It remains to
evaluate the currents $j(0,1,0)$ and $j(1,1,0)$.

Consider first $j(0,1,0)$, the rate to overcome the barrier between wells $i$ and $i+1$,
given that both $i-1$ and $i+1$ are empty.  The mean first-passage
time $\tau_{ca}$, for a particle to overcome the barrier,
is readily found via analysis of the one-dimensional
Fokker-Planck equation.  From the standard result \cite{VanKampen} we have

\begin{equation}\label{FPT}
 \tau_{ca} = \frac{1}{k_BT}\int_a^ce^{U(x')/k_BT}dx'\int_{-\infty}^{x'}e^{-U(x'')/k_BT}dx'',
\end{equation}

\noindent where $a=x_0$ and $c=x_0 + 2d_0$ are the positions of adjacent potential minima,
with $x_0$ given by Eq.~(\ref{x0}).
Using the effective external potential, Eq.~(\ref{PotExtEfe}), for $U(x)$
(since there are no interactions with other particles),
numerical evaluation
of Eq.~(\ref{FPT}) yields $\tau_{ca} = 6.538\,$s$\, = 1/j(0,1,0)$. This value agrees to within uncertainty
with our simulation result  for the mean time $\tau$ for a particle to hop between adjacent wells
when there are no other particles in the system.  Thus the mean-field curve in
Fig.~\ref{J_d} agrees with simulation in the low-density limit.

To estimate the transition rate $j(1,1,0)$,
we perform a Monte Carlo simulation to determine the mean time required for a particle
to hop to the next well, when the preceding
well is occupied, and there are no other particles in the system.  This yields
$1/j(1,1,0) = 5.548\,$s.  Although the presence of the trailing particle leads
to an increase of about 18\% in $j$, the effect is not sufficient to yield
quantitative agreement with the current observed at higher densities (Fig.~\ref{J_d}).
There are two possible sources for this
discrepancy.  First, at moderate and high densities, strings of $n \geq 3$ occupied wells
occur with finite probability, and
the cumulative effect of repulsions along the chain should make the hopping rate of the
lead particle an increasing function of $n$.
Simulations of $n = 3$ - 9 occupied wells show that the transition rate of the first
particle grows with $n$, but not enough to account for the maximum value of the current observed.

A second point is that the mean-field factorizations
$P(0,1,0) = \rho(1-\rho)^2$ and $P(1,1,0) = \rho^2(1-\rho)$,
are not very accurate in the continuous-space model.  For density $\rho \simeq 0.52$, for example,
we find

\begin{equation}
\frac{P(0,1,0)}{\rho(1-\rho)^2} \simeq 1.21
\end{equation}
and
\begin{equation}
\frac{P(0,1,1)}{\rho^2(1-\rho)} \simeq 1.04
\end{equation}
implying a significant correction to the mean-field theory predictions.

\subsection{Fluctuations}

\begin{figure}[htb]
\begin{center}
\includegraphics[height=6cm]{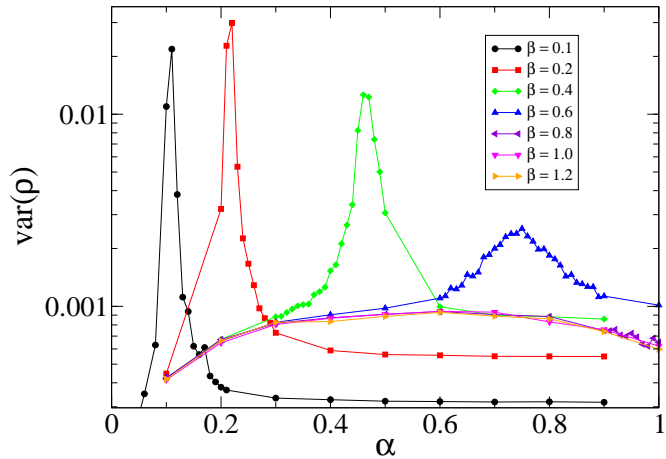}
\end{center}
\caption{Density variance versus $\alpha$ for $\beta$ values as indicated. System size $L=200$.}
\label{var_den}
\end{figure}

To close this section we note an
interesting finding on fluctuations. The variance of the density,
as a function of $\alpha$, with
fixed $\beta$, exhibits a maximum at the discontinuous transition
(see Fig.~\ref{var_den}).  For $\beta$ values such that the transition is continuous,
by contrast, no peak in var($\rho)$ is observed at the transition.  (Similar
behavior is found, varying $\beta$ with $\alpha$ fixed.)  The large density fluctuations
are associated with the presence of a shock separating high- and low-density regions, whose
position fluctuates over the entire system.
The position of maximum variance agrees to within uncertainty with the
lines of the (discontinuous) phase transitions reported in Fig.~\ref{PhaDia}.
Analysis of the variance, however, appears to furnish less precise
results than the method described above.
The total energy exhibits fluctuations similar to those observed in the density,
but we do not find any signal in var($J$) associated with the phase transitions.

\vspace{1cm}

\vspace{1em}

\section{Discussion}\label{CON}

We propose a continuous-space model of interacting Brownian particles in a periodic
potential as a possible realization of the TASEP.
The particles are
subject to a constant drive in a periodic external potential.
Using numerical integration of the Langevin equation and Monte Carlo simulation,
we study systems of L = 100, 200, and 500 wells.
Our results show that the continuous-space model exhibits continuous and discontinuous
phase transitions analogous to those observed in the lattice TASEP. The phase diagram of the
continuous-space model is similar to that of the lattice model, but exhibits some differences
due to the absence of particle-hole symmetry. This difference appears to be
associated with fluctuations of particle positions around potential minima.
Such fluctuations, together with the repulsive interactions
between neighboring particles, cause the current to attain its maximum value at a density
somewhat greater than 1/2, the density marking the maximum current in the lattice model.
We expect these changes (relative to the lattice model) to be generic for
continuous-space systems exhibiting TASEP-like phase transitions.

We believe that the present study demonstrates the possibility of observing TASEP-like
behavior in laboratory experiments on systems of interacting colloidal particles in a one-dimensional
optical tweezer array.  The essential features of the TASEP - localization of particles in
potential wells, with multiple occupancy prohibited, and bias hopping along the line - are
readily accomplished with a appropriate choice of particle, fluid, and tweezer array parameters.
It is however less obvious how to implement random insertion and removal of particles at the
first and last wells of the array.  Particle manipulation can be accomplished using
optical tweezers to transfer particles between wells and reservoirs.
To transfer particles in a random fashion, these tweezers would have to be intrinsically noisy
or chaotic,
controlled by a random number generator, or driven by a noise signal.
A simpler alternative may be {\it periodic} insertion and removal.  In this case, one
inserts a particle into the first well (when empty) at intervals of $\tau/\alpha$, and
checks for occupancy of the final well at intervals of $\tau/\beta$, removing the particle
if the well is occupied.

\begin{figure}[htb]
\begin{center}
\includegraphics[height=7.5cm]{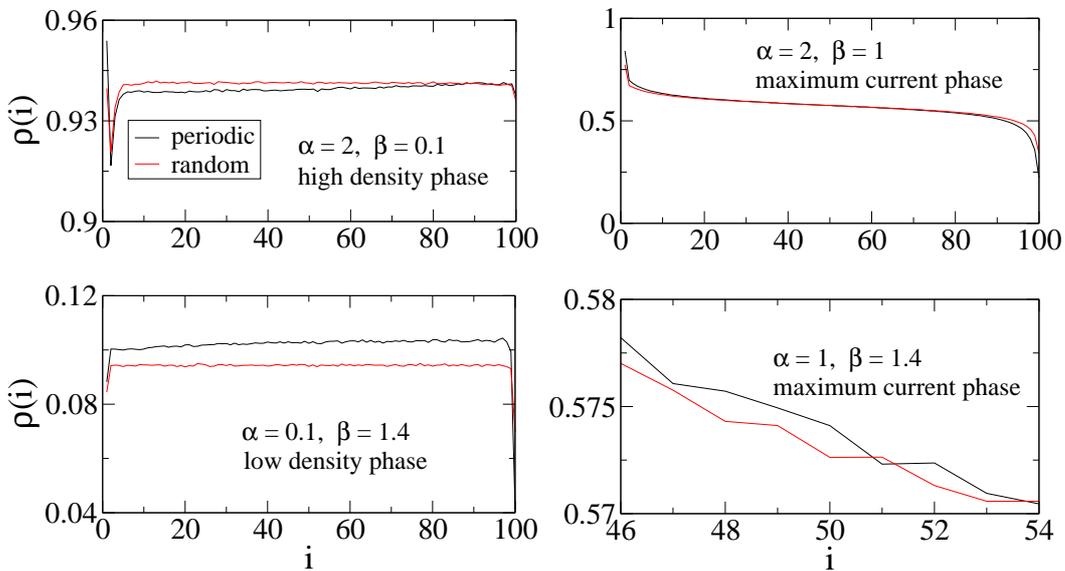}
\end{center}
\caption{Comparison of density profiles obtained using periodic (black) and random (red)
insertion and removal, $L=100$.}
\label{perfis_periodicos}
\end{figure}

A preliminary study of the continuous-space model using periodic insertion
and removal confirms that the three
TASEP phases are again found.  The density profiles
under periodic and random particle transfer are very similar in the maximum current phase, where the
density profile is insensitive to small changes in the insertion and removal rates.
Small systematic differences do however appear in the other phases, as shown in Fig.~\ref{perfis_periodicos}.

Although our study strongly suggests the feasibility of a laboratory realization of the TASEP,
a number of additional features would have to be included in the model, before a quantitative
comparison with experiment could be made.  The principal modifications we expect to be necessary
are study of a three-dimensional model, allowing fluctuations in directions perpendicular to the
array axis, and inclusion of hydrodynamic interactions between the particles.  We defer these
tasks to future work.

\begin{acknowledgments}
We thank O. N. Mesquita, U. Agero and J. G. Moreira for helpful comments.
This work was supported by CNPq and Fapemig, Brazil.
\end{acknowledgments}

\end{document}